\documentstyle[aps,preprint]{revtex}
\tighten
\begin{document}
\draft 
\preprint{July 30, 1999}
\title
{
Transitions 
from the Quantum Hall State
to the Anderson Insulator:
Fate of Delocalized States
}
\author{
Y. Morita$^1$,
K. Ishibashi$^1$ and
Y. Hatsugai$^{1,2}$
}
\date {July 30, 1999}
\address
{
Department of Applied Physics, University of Tokyo,
7-3-1 Hongo Bunkyo-ku, Tokyo 113, Japan$^1$ \\
PRESTO, Japan Science and Technology Corporation$^{2}$
}
\maketitle
\begin{abstract}
Transitions between 
the quantum Hall state and 
the Anderson insulator
are studied in a two dimensional 
tight binding model
with a uniform magnetic field 
and a random potential.
By the string (anyon) gauge,
the weak magnetic field regime
is explored numerically.
The regime is closely related to the continuum model.
The change of the Hall conductance and
the trajectoy of the delocalized states
are investigated by 
the topological arguments and the Thouless number study.
\end{abstract}

\pacs{73.40.Hm,73.50.-h}

\narrowtext
\section{Introduction}
In the integer quantum Hall effect,
delocalized states near the center of each Landau band
play a crucial role\cite{iqhe}.
On the other hand,
when 
the randomness strength becomes 
as large as the energy cutoff,
it is expected that
all the states become localized\cite{aalr}.
An appealing scenario for the transitions,
which is called the 'floating' or 'levitaion' of
the delocalized states, was proposed in refs.\cite{fl1,fl2,fl3}.
It claims that
the delocalized states do not diappear discontinuously
but float upward in energy
with the increase of the randomness strength.
Finally,
when all the delocalized states rise above the Fermi energy, 
the system becomes the usual Anderson insulator.
Then
a crossover occurs
from the unitary class to the orthogonal class.
Furthermore,
based on the scenario, the global phase diagram was proposed
for transitions 
between different quantum Hall states\cite{klz}.
Recently 
some controversies arise
about the trajectory of the delocalized states 
\cite{tbm2,tbm1,tbm3,him,tbm4}.
There are also experimental discussions on 
the validity of the global phase diagram 
\cite{bd1a,bd1b,bd1c,bd1d}.

In this paper,
we study
transitions between 
the quantum Hall state and the Anderson insulator
in a tight binding model.
The weak magntic field regime
is especially focused on.
It is closely related to the continuum model.
The topological invariant 
for each energy band (Chern number)
and the Thouless number are used 
as probes for the transitions.
The summation of the Chern number 
below the Fermi energy
gives the Hall conductance.
The Thouless number tells us the location 
of the delocalized states.
This paper is an extended version of our previous letter
\cite{him} including new data and more detailed discussions.

The present article is organized as follows.
In the next section,
the tight binding Hamiltonian is defined and
the string (anyon) gauge is introduced.
The string gauge is essential 
for the numerical realizaion of 
the weak magnetic field regime.
In section III, 
key observables in our study
are defined:
the topological invariant for each energy band (Chern number) 
and the Thouless number.
Section IV is devoted to the statement of
a sum rule in the change 
of the Hall conductance.
In Section V,
the weak magnetic field regime is studied
in connection with the continuum model.
The breakdown of the quantized Hall conductance
due to randomness is demonstrated and
the trajectory of the delocalized states is investigated.
\section{Model}
\subsection{Hamiltonian}
The tight binding Hamiltonian is
\begin{eqnarray}
{\cal H}
=
{\sum _{{\langle}i,j{\rangle}}}
c_{i}^{\dagger}t_{ij}c_{j}+h.c.+
{\sum _{i}}
w_{i}c_{i}^{\dagger}c_{i},
\nonumber
\end{eqnarray}
where $t_{ij}=\exp(ia_{ij})$,
$a_{ij}$ denotes a gauge potential and
${{\langle}i,j{\rangle}}$ refers to a nearest neighbor.
A magnetic field per plaquette
is given by 
$\phi
={\displaystyle \sum_{\put(0,0){\framebox(4,4)}}}
{a_{ij}}/2{\pi}=p/q$
($p$ and $q$ are coprime integers)
where the summation runs over four links around a plaquette.
The operator $c_{i}^{\dagger} (c_{i})$ 
creats (annihilates) a spinless fermion
at a site $i$.
A random potential at a site $i$ is expressed by  $w_{i}=Wf_{i}$
where $f_{i}$'s are uniform random numbers and chosen from $[-1/2,1/2]$.
See, for example, refs.\cite{tbm2,tbm1,tbm3,ando1,mk1}
for previous works on this model.
\subsection{String (Anyon) gauge}
The integer quantum Hall effect 
is believed to be described by the continuum model.
As will be discussed later,
the weak magnetic field regime
in the tight binding model
is closely related to it.
Therefore, in this paper, 
we focus on the weak magnetic field regime 
(e.g. $\phi=p/q=1/64$).
In order to realize the regime numerically,
the choice of a gauge potential
plays an important role.
Here we employ the string (anyon) gauge\cite{him}.
An example of the string gauge
is shown in Fig.1.
The extension to other geometries 
is straightforward.
Choosing a plaquette $S$ as a starting one,
we draw
outgoing arrows (strings)
from the plaquette $S$.
$a_{ij}$  on a link $ij$
is given by $\phi n_{ij}$ where
$n_{ij}$ is the number of strings
cutting the link $ij$
(the orientation is taken into account).
In the string gauge,
the compatible magnetic field $\phi$
with the periodicity of the system $L_{x}{\times}L_{y}$
is 
$\phi={n}/{L_{x}L_{y}},{\;}n=1,2,{\cdots},L_{x}L_{y}$
(on the other hand, 
$\phi={n}/{L_{x(y)}},{\;}n=1,2,{\cdots},L_{x(y)}$ 
in the Landau gauge).
Therefore, for a square $L\times L$ geometry, 
one is able to use
$L$ times smaller magnetic field 
in the string gauge
than in the Landau gauge.
\section{Observables}
In this section,
we define
key observables
to study transitions
between the quantum Hall state 
and the Anderson insulator.
One is the topological invariant 
for each energy band ( the Chern number).
It is closely related to the Hall conductance.
The other is the Thouless number. 
It tells us how the wavefunctions are extended spatially.
\subsection{Topological invariant (Chern number)}
The topological invariant for each energy band,
Chern number,
plays an important role in the following arguments.
Here the periodicity of $L_{x}{\times}L_{y}$
is imposed on $a_{ij}$ and $w_{i}$
and the momentum ${\bf{k}}$ is well-defined
(the infinite size limit corresponds to
$L_{x},L_{y}{\rightarrow}\infty$).
The Chern number for the $n$-th band, $C_n$,
is defined by
\begin{eqnarray}
 C_n =\frac 1 {2\pi i} 
\int d{\bf k}\; \hat z \cdot ({\bf{\nabla}}&\times&{\bf A}_n),\;  
 {\bf A}_n = \langle u_n ({\bf{k}}) | {\bf{\nabla}} | u_n({\bf{k}})
 \rangle,
\nonumber
\end{eqnarray}
where ${\nabla}={\partial}/{\partial}{\bf k}$,
$| u_n(\bf{k}) \rangle$ is a Bloch wavefunction of the 
$n$-th band with $L_xL_y$ components
and the $\gamma$-th component is $u_n^\gamma(\bf{k})$.
The integration $\int d{\bf k}$ runs over the Brillouin zone.
When 
the Fermi energy lies 
in the lowest $j$-th energy gap,
the Hall conductance $\sigma_{xy}$
is given by 
$\sigma_{xy}=
{\displaystyle \sum_{n=1}^j} C_n$\cite{tknn1,tknn2}.
When all states in a band are localized,
the Chern number vanishes.
On the other hand,
states with a non-zero Chern number
contribute to the $\sigma_{xy}$ and we call them 'extended states'.

Arbitrarily 
choosing the $\alpha$ and $\beta$-th components of the wavefunction, 
another expression of the Chern number is written as
\begin{eqnarray}
C_n=\sum_\ell N_{n\ell},\;\;\; N_{n\ell}=\frac 1 {2\pi} 
\oint_{\partial R_\ell}d{\bf{k}} \cdot {\bf{\nabla}} {\rm Im\  ln\ } 
(\frac {u^\alpha_n({\bf{k}})}  {u^\beta_n({\bf{k}})}).
\nonumber
\end{eqnarray}
Here
$\bf{k}_\ell$
is a zero point (vortex)
of $u_n^{\alpha}(\bf{k})$ in the Brillouin zone,
$N_{n\ell}$ is
a winding number (charge)
of the vortex,
$R_\ell$ is a region around $\bf{k}_\ell$ 
which does not include either zero
of the $\alpha$-th or $\beta$-th component,
and $\partial R_\ell$ is the boundary.
The arbitrariness 
in the choice of indices $\alpha$ and $\beta$
is a kind of gauge degree of freedom.
The configuration of the vortices
depends on the gauge choice but
the physical observables, 
e.g. the Hall conductance and
the position of each extended state, 
are gauge invariant.
\subsection{Thouless number}
The Thouless number $g(E)$ is widely used
in the study of the Anderson localization
\cite{tn1}
(see also ref.\cite{ando1}).
It tells us 
how the wavefunctions at energy $E$
are extended spatially.
Here a finite $L{\times}L$ cluster is used
with a periodic 
or an antiperiodic
boundary condition.
It
is defined by 
\begin{eqnarray}
g(E)=v(E)/w(E)
\nonumber
\end{eqnarray}
where the $v(E)$ is an energy shift 
by replacing the boundary condition
and $w(E)$ is a local level spacing.
By fitting the $g(E)$'s 
to the form $g(E)=g_{0}{\exp}(-L/\xi(E))$,
one can determine the $\xi(E)$
which is the localization length
when $L$ is sufficiently large.
However,
when the condition $L>\xi(E)$ is not satisfied,
the $\xi(E)$ can not 
be identified with the localization length.
For example,
in the {\it intermediate region} discussed below,
the localization length is extremely long and
beyond our available system size.
\section{change of the Hall conductance
and sum rule}
Before going to 
the weak magnetic field regime,
let us explain 
the change of ${\sigma}_{xy}$ due to randomness and 
a {\it sum rule} in the transitions
\cite{him,sum,drc2,drc1}.
By fixing the 'gauge'
in the definition of the Chern number,
the configuration of vortices are determined
in each energy band.
Results for
$\phi=p/q=1/3$ and $2/5$ 
are shown as examples (Figs.2).
As the randomness strength $W$ is increased,
vortices in each band 
move in energy.
The motion of each vortex
is continuous as a function of $W$
and it forms a vortex line.
Therefore the only way for vortices to appear or vanish
is the pair-creation or pair-annihilation 
of them with opposite charges.
Although vortices move with the change of $W$,
the Chern number does not change generally
due to its topological stability.
When the two bands touch at a critical value
of $W=W_{0}$,
the spectrum generally becomes that of the Dirac fermion
\cite{dirac1,dirac2,dirac3}.
Define $(k_{x}^{0},k_{y}^{0})$ by the gap-closing point
in the Brillouin zone
and focus on the neighborhood i.e.
${\bf{p}}=
{}^{t}\pmatrix{k_{x}-k_{x}^{0}, k_{y}-k_{y}^{0}, W-W_{0}}
\sim{\bf 0}$.
Then the leading part of the Hamiltonian is, in generic,
given by
\begin{eqnarray}
H_{0}({\bf p})
=
1{\bf v}_{0}{\bf p}
+
({\sigma}_{x},{\sigma}_{y},{\sigma}_{z})
\bf{v}
\bf{p},
\nonumber
\end{eqnarray}
where 
${\bf v}_{0}$ is a $1{\times}3$ vector,
${\sigma}_{x(y,z)}$ is a  $2{\times}2$ Pauli matrix 
and ${\bf v}$ is a $3{\times}3$ matrix.
By performing the unitary transformation and
the redefinition of ${\bf p}$,
the $H_{0}$ reduces to \cite{drc2}
\begin{eqnarray}
H_{1}({\bf p})
&=&
1{\bf v}_{0}{\bf p}+
{\sigma}_{x}p_{x}+
{\sigma}_{y}p_{y}+
{\sigma}_{z}p_{z}{\rm sgn}({\rm det}({\bf v})).
\nonumber
\end{eqnarray}
By diagonalizing the $H_{1}$,
it can be seen that
a vortex moves from one band to the other
at the gap-closing point
and the Chern number for each band changes.
Then the continuity requires that
the total Chern numbers of the two bands
do not change.
This is the sum rule \cite{him,sum,drc2}.
Moreover,
the change of $\sigma_{xy}$
with a fixed electron density
generally obeys the selection rule
$\Delta\sigma_{xy}
={\rm sgn}({\rm det}{\bf v})
={\pm}1$\cite{klz,him,drc2}.
However, 
the change of 
the {\it observed} Hall conductance $\sigma_{xy}^{phys}$
can break the selection rule (anomalous plateau transition).
Here the definition of $\sigma_{xy}^{phys}$ is
\begin{eqnarray}
\sigma_{xy}^{phys}
=\lim _{T\to 0}\lim _{L\to \infty}
\sum_{n}
f(E_{n}){\ }C_{n}
\nonumber
\end{eqnarray}
where
$E_{n}=E_{n}({\bf k}={\bf 0})$ and
$f(E)$ is the Fermi distribution function.
Then 
the $\sigma_{xy}^{phys}(L)$ 
is obtained as
the averaged $\sigma_{xy}(L)$
within an energy window
$|E-E_{f}|{\sim}{\cal O}(1/L)$.
Therefore
we assume 
\begin{eqnarray}
\sigma_{xy}^{phys}(L){\cong}\overline{\sigma}_{xy}(L)
\nonumber
\end{eqnarray}
where $L$ is sufficiently large and
$\overline{\sigma}_{xy}(L)$ is the averaged $\sigma_{xy}(L)$
over different realizations of randomness \cite{aa}. 
The $\sigma_{xy}(L)$ is quantized in each realization of randomness
even when the Fermi energy lies in a small gap ${\sim}{\cal O}(1/L)$.
However,
since the small gap
is sensitive to the randomness realization,
the quantization
breaks down after the ensemble average.
Therefore
the anomalous plateau transition 
$\overline{\sigma}_{xy}=n{\rightarrow}0$ occurs
(see also ref.\cite{him}).
It is in contrast to
the plateau transition 
due to the change of
electron density where
$\Delta\sigma_{xy}=\pm 1$ holds
even after the ensemble average.
\section{Weak magnetic field regime}
Now we shall explore 
the weak magnetic field regime,
which is closely related to
the continuuum model.
In the following,
$\phi$ is fixed as
$\phi=p/q=1/64$.
It is realized numerically
by the string gauge discussed above.
Transitions between
the quantum Hall state and the Anderson insulator
are investigated
by the topological invariant 
for each energy band (Chern number)
and the Thouless number.
The summation of the Chern number 
below the Fermi energy
gives the Hall conductance.
The Thouless number tells us the location 
of the delocalized states.
\subsection{Breakdown of quantum Hall states}
In Figs.3,
the density of states (DOS)
and the Thouless number $g(E)$
are shown for different randomness strengths $W$'s.
In Fig.4, 
the voritces are shown 
as a function of $W$.
In Figs.5, 
the averaged local Chern number
$\overline{C(E)}$ (see \cite{apx1} for a precise definition)
and its variance $(\overline{{\delta}C(E)^{2}})^{1/2}$
are shown with different $W$'s.
Here
$\overline{\cal O}$ denotes
an averaged observable $\cal O$
over different realizations of randomness.
Moreover, in Figs.6,
the Chern number
of the $i$-th band $C_{i}$
and $\overline{C_{i}}$
are shown 
as a function of $W$.

Let us first 
investigate a region
with sufficiently weak randomness.
As seen in Fig.3 (a),
there are well-defined Landau bands.
The Thouless number $g(E)$ shows a sharp peak 
near the center of each Landau band.
It is consistent with the result that
all the states are localized
except at a single energy in each Landau band\cite{ando1}.
The number of states extended over a $L{\times}L$ sample
is ${\sim}L^{2-1/{\nu}}$ and 
is not macroscopic.
The Chern number in each Landau band
is $+1$ except in the center band.
The sum of all Chern numbers is always zero and
the center band has negative Chern number $-62$.
When the Fermi energy lies in the Landau gap
or near the band edge, 
the $\sigma_{xy}$ does not
depend on the randomness realization generally.
Then
the $\sigma_{xy}$ is quantized to an integer
even after the ensemble average
and the system belongs to the quantum Hall states.

Next let us consider the strong-randomness effects.
With the increase of $W$,
the Landau gaps collapse 
from the center ($E=0$) to the bottom (see Figs.3).
As seen in Figs.4-6,
it is associated with 
many pair-creations and -annihilations.
Extended states with a negative Chern number
spread from the center to the bottom 
through the pair-creations and -annihilations and
an energy region
where
$\overline{C(E)}{\sim }0$ but
$(\overline{{\delta}C(E)^{2}})^{1/2}{\neq}0$ 
extends over the whole spectrum
(see Figs.5 (c) and (d)).
We call it the {\it intermediate region}.
In this region,
the mean free path $\ell$ is comparable with 
the magnetic length $ \ell_B(\sim 1/ \sqrt \phi)$ and
the internal structure within the magnetic unit cell is irrelevant.
Then,
since the magnetic unit cell contains multiple of the flux quanta,
it is effectively gauge equivalent to the zero magnetic field and
a crossover occurs 
from the unitary class to the orthogonal class. 
In the region,
the Chern number in each band
is generally non-zero and quantized to an integer
for a given realization of randomness.
However, the Chern number
depends strongly on the randomness realization.
Then the averaged Chern number is not an integer
and takes a small value
due to the cancellation of
positive and negative Chern numbers
in different realizations.
Therefore, after the ensemble average,
the $\sigma_{xy}$ is no longer quantized to an integer.
In the region,
the energy position
of each extended state
depends strongly on the randomness realization.
The number of the extended states is not macroscopic and
we speculate that
the probability measure 
for the delocalized states to exist is zero
in the infinite size limit
(similar region was found in the random magnetic field problem
\cite{rfx1,rfx2}).
When the randomness is stronger,
the extended states no longer exisit and 
$C(E)=0$ 
in {\it each} realization of randomness.
Then the system belongs to the usual Anderson insulator
and the $\sigma_{xy}$ becomes zero.

To summarize:
for sufficiently weak randomness, 
there are delocalized states in each Landau band
and the system is in the quantum Hall state.
As the randomness strength is increased,
the Landau gaps collapse from the center ($E=0$) to the bottom
and the intermediate region
extends over the whole spectrum.
When the Fermi energy lies in the region,
the $\sigma_{xy}$ is no longer quantized to an integer.
With 
sufficiently strong randomness,
the system change into 
the usual Anderson insulator and the $\sigma_{xy}=0$.
\subsection{Scaling in the weak magnetic-field limit}
In the above subsection,
the magnetic-field strength is kept as $\phi=p/q=1/64$
and the randomness strength $W$ is varied.
In this subsection,
we comment on the scaling 
in the weak magnetic-field limit 
($\phi=1/q{\ }{\rightarrow}{\ }0$).

In the limit ( $\phi{\ }{\rightarrow}{\ }0$ )
without randomness $W=0$,
the rescaled energy spectrum 
$E/E_{g}$
becomes the Landau levels
and the system reduces to the continuum model
($E_{g}{\ }{\sim}{\ }\Gamma _{c}/q$ is the lowest band gap
and ${\Gamma}_{c}(\sim 1)$ is the energy cut-off in the lattice model).
In the limit,
the magnetic length $l_{B}$ is much larger 
than the lattice spacing i.e.
$l_{B}=1/\sqrt{\phi}=\sqrt{q}{\ }{\rightarrow}{\ }\infty$.

When the $W$ is increased to ${\sim}W_{c}$,
the width of the lowest Landau band becomes 
comparable with the Landau gap.
In the weak field limit,
$W_{c}/{\Gamma}_{c}{\ }{\rightarrow}{\ }0$ 
but  $W_{c}/E_{g}{\ }{\rightarrow}{\ }\infty$\cite{tbm3}.
It implies that 
the lowest Landau band
is robust in the continuum limit.
Therefore, before the lowest Landau gap closes,
higher Landau bands merge and 
the reconstruction of eigenstates occurs
through the pair-creation and -annihilation.
Then the 'floating' is not the only way 
for the delocalized states to disappear.
For stronger randomness
(${\Gamma}_{c}{\ }{\gtrsim}{\ }W{\gtrsim}{\ }W_{c}{\ }$),
all the states are in the intermediate region.
In this region,
the localization length is extremely large.
Finally, when $W{\ }{\gtrsim}{\ }{\Gamma}_{c}$, the system 
belongs to the usual Anderson insulator.
\subsection{Fate of delocalized states}
At last,
let us discuss 
the fate of delocalized states.
In Fig.7,
the trajectory of the delocalized states
is shown 
for
the lowest Landau band ($n=0$)
and the second one ($n=1$).
In order to search the trajectory,
the Thouless number $g(E)$ is obtained
for
$L_{x}{\times}L_{y}=24{\times}24$,
$32{\times}32$ and $40{\times}40$
(the ensemble average is performed 
over $100$ different realizations of randomness).
The energy position of the delocalized states 
is obtained
by searching 
the peak position of $g(E)$ in each Landau band.
For the full circles,
we also confirmed that
the localiztion length $\xi(E)$ is 
the largest at the point in each Landau band
(see the inset of Fig.7 as an example).
It can be observed that
their positions 
are going down in energy
with the increase of randomness strength $W$.
At the same time, as seen in Figs.3,
the density of states also shifts downwards.
Then, in the lowest Landau band, 
the delocalized states 
'floats' or 'levitates'
slightly\cite{hal}.
However there is no sign of the floating
across  
the  Landau gap.
The way how the delocalized states disappear
is also shown in Fig.7.
After the collapse of the Landau gap,
the intermediate region extends over the whole spectrum and
the delocalized states mix in the region.
Even after the collapse,
the peak positions of $g(E)$ can be defined for each Landau band 
and they are also shown by the open circles.
In the region, the positions have a large fluctuation
and they finally merge when $W$ is sufficiently large.
In fact, 
for a given realization of randomness,
we can further trace 
the trajectory of states 
which carry non-zero Chern number,
and it finally disappear 
through the pair-annihilation\cite{him}.
However, since
their positions in the region
depend strongly on the randomness realization,
we speculate that
the probability density  
for the states to be delocalized is zero.
\section{Summary}
In summary,
we studied
transitions
between 
the quantum Hall state and 
the Anderson insulator
based on the tight-binding model.
Aspects of the transitions
are revealed 
by the topological arguments 
and the Thouless number study.
In the weak randomness region,
there are delocalized states
in each Landau band and 
the system is in the quantum Hall state.
As the randomness becomes stronger,
the Landau gaps collapse 
from the center ($E=0$) to the bottom.
Accompanied by the collapse,
the intermediate region 
with a large localization length
extends over the whole spectrum. 
When the Fermi energy lies in the region,
the Hall conductance 
is not generally quantized to an integer.
The delocalized states in each Landau band
mix in the region successively and, finally,
the system belongs to the Anderson insulator.

This work was supported in part by Grant-in-Aid
from the Ministry of Education, Science and Culture
of Japan and also Kawakami Memorial Foundation.
The computation  has been partly done
using the facilities of the Supercomputer Center,
ISSP, University of Tokyo.

\begin{figure}

\caption{
The definition of the string (anyon) gauge
for a $3{\times}3$ square system
with periodic boundary condition.}

\caption{
Zero points (vortices) of the Bloch wavefunctions
with their winding numbers (charges).
The shaded regions show energy bands.
Each circle denotes 
the energy position of a vortex 
as a function of randomness strength $W$.
The full circle means the charge $+1$ and the open circle $-1$.
When the Fermi energy lies in the energy gap,
the $\sigma_{xy}$ is quantized to an integer.
The integer is also shown in each energy gap.
(a) $\phi=1/3$ and $L_{x}{\times}L_{y}=3{\times}3$ and
(b) $\phi=2/5$ and $L_{x}{\times}L_{y}=5{\times}1$.}

\caption{
The density of states (DOS)
and the Thouless number $g(E)$
for $\phi=1/64$ and $L_{x}{\times}L_{y}=40{\times}40$,
where the ensemble average
is performed 
over $100$ different realizations of randomness.
(a) $W=0.346$, (b) $W=1.386$ and (c) $W=5.196$.}

\caption{Zero points (vortices) of the Bloch wave functions
with their winding numbers (charges) 
for $\phi=1/64$ and $L_{x}{\times}L_{y}=24{\times}24$
(see also Figs.2).}

\caption{
The averaged local Chern number 
$\overline{C(E)}$ (the solid line)
and its variance 
$(\overline{{\delta}C(E)^{2}})^{1/2}$ (the broken line)
for $\phi=1/64$ and $L_{x}{\times}L_{y}=8{\times}8$,
where the ensemble average
is performed 
over $100$ different realizations of randomness.
(a)$W=0.476$, (b)$W=1.169$, (c)$W=2.641$ and (d)$W=6.495$.}

\caption{
The Chern number of the $i$-th band $C_{i}$
for $\phi=1/64$ and $L_{x}{\times}L_{y}=8{\times}8$
as a function of randomness strength $W$.
It is to be noted that
$C_{i}=+1$ except at the center band
for sufficiently weak randomness.
(a) the realization of randomness is fixed. 
(b) the ensemble average is performed 
over $100$ different realizations.}

\caption{
The trajectory of delocalized states
for $\phi=1/64$ 
is denoted by the full circles. 
For the open circles,
although the Thouless number $g(E)$ has a peak at the point,
the corresponding states belong to the intermediate region.
The inset shows the density of states (DOS)
and the inverse localization length $1/{\xi(E)}$
for $W=1.039$.
The ${\xi(E)}$ is obtained 
by fitting the $g(E)$'s
to the form $g_{0}{\exp}(-L/\xi(E))$.}

\end{figure}


\begin{references}
\bibitem{apx1}
The local Chern number $C(E)$ is defined by
$C(E)=\sum_{i}C_{i}f_{i}(E)$.
Here $f_{i}(E)$ is defined as follows:
$f_{i}(E)=1$ when $E$ belongs to the $i$-th band
and, otherwise, $f_{i}(E)=0$.
\end{references}
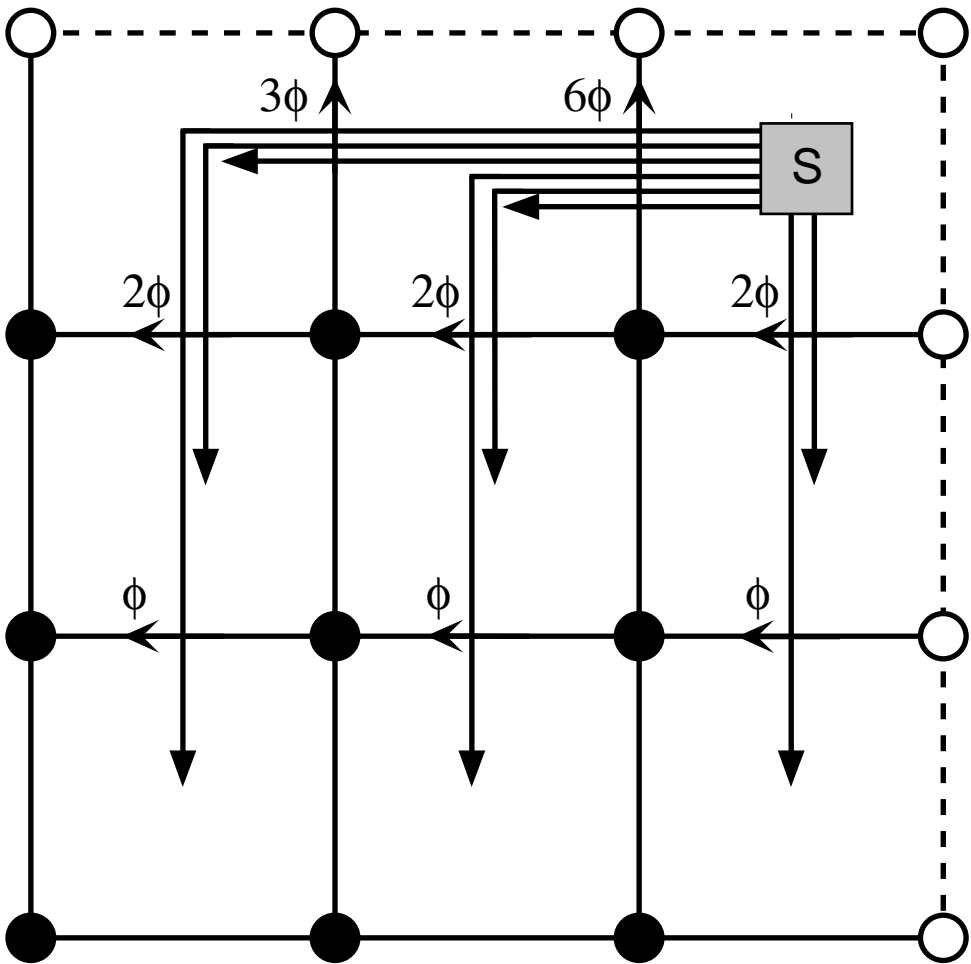
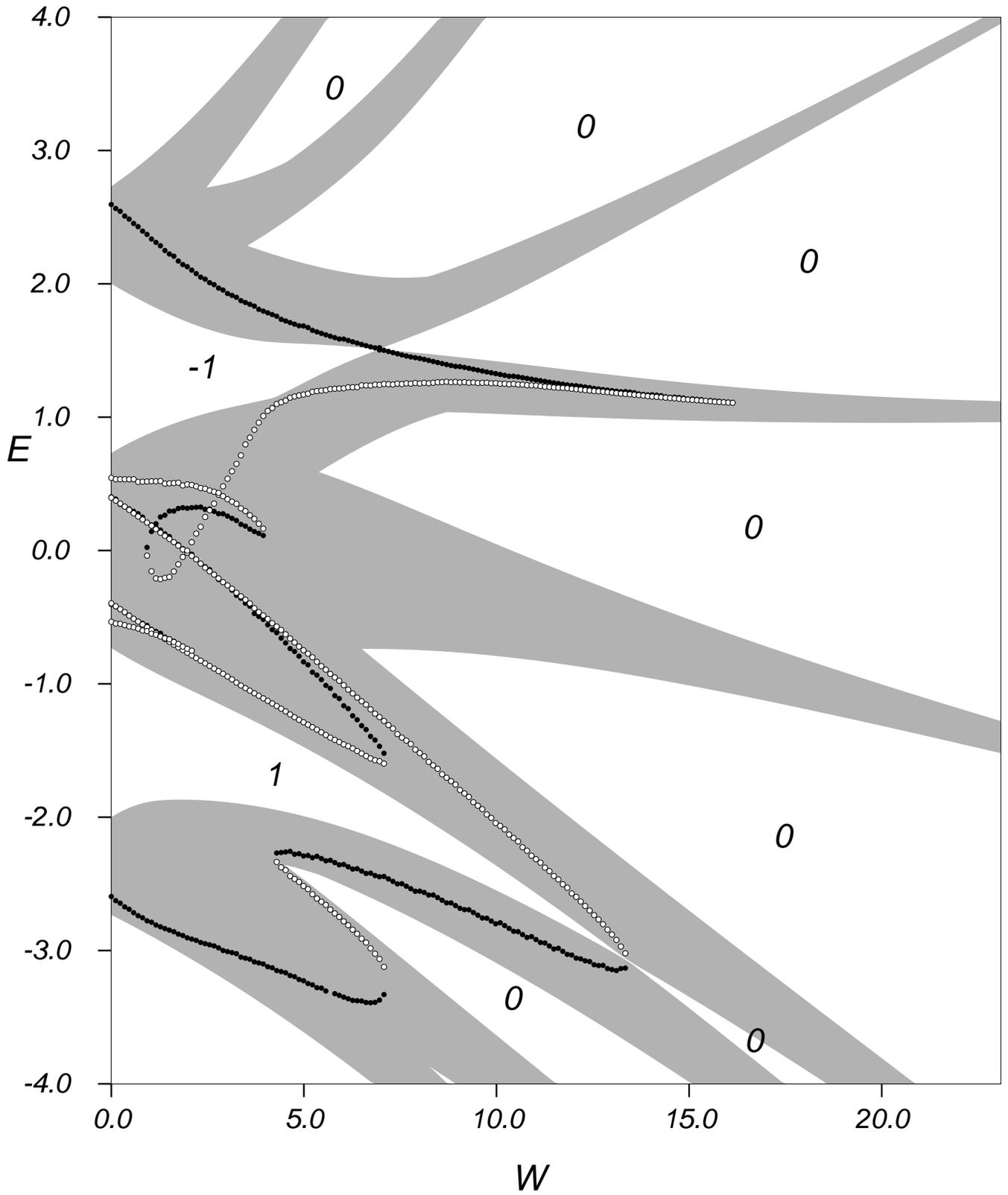
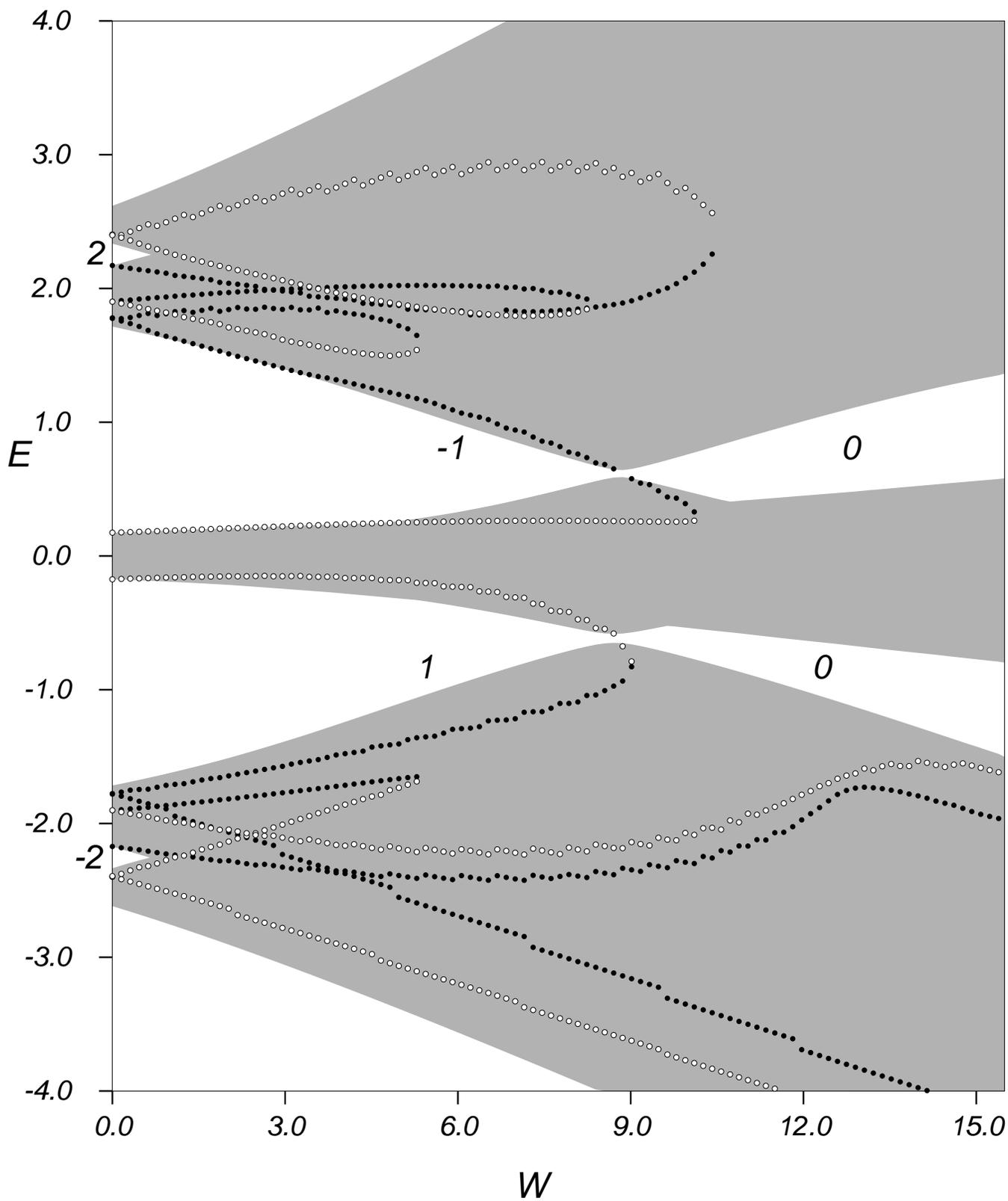
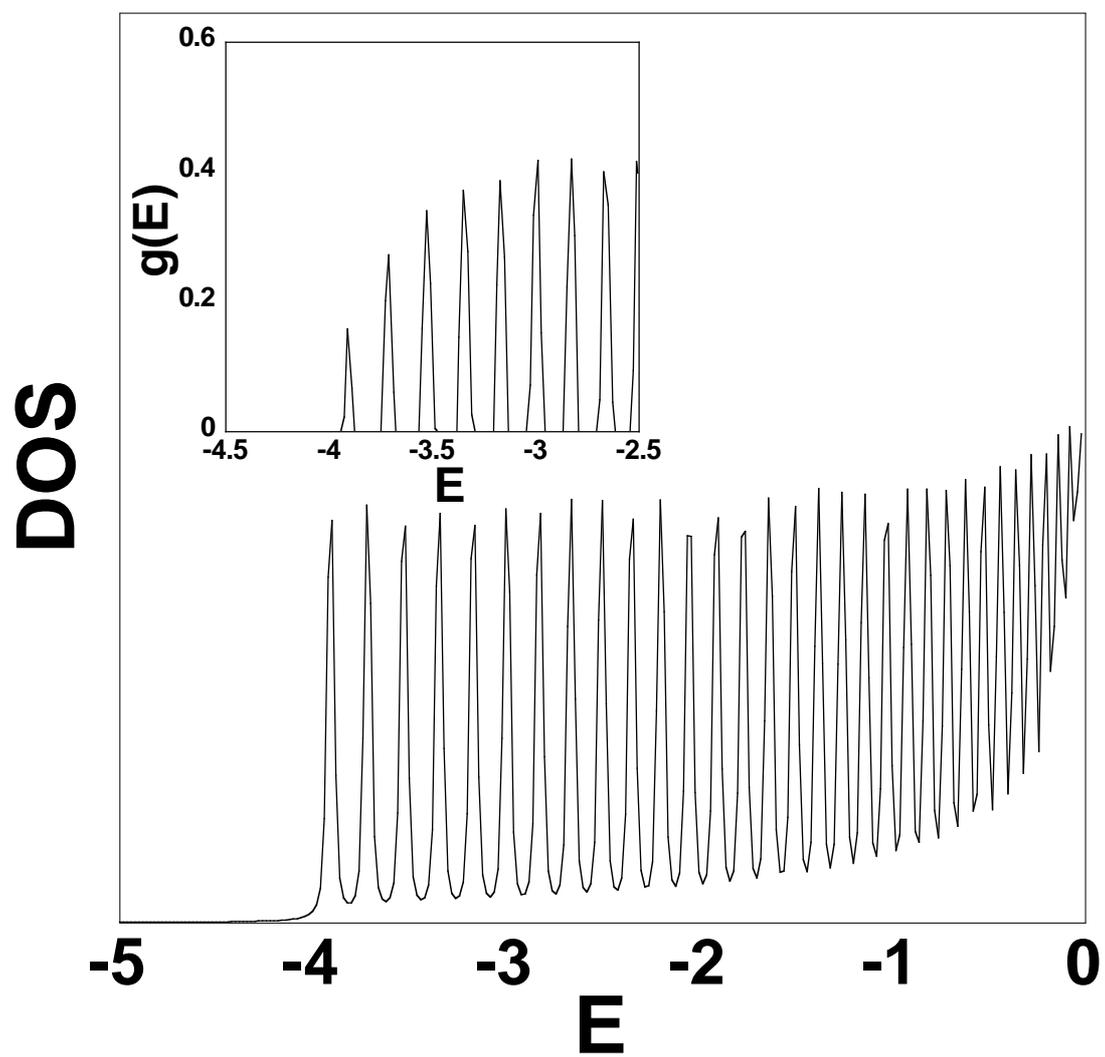
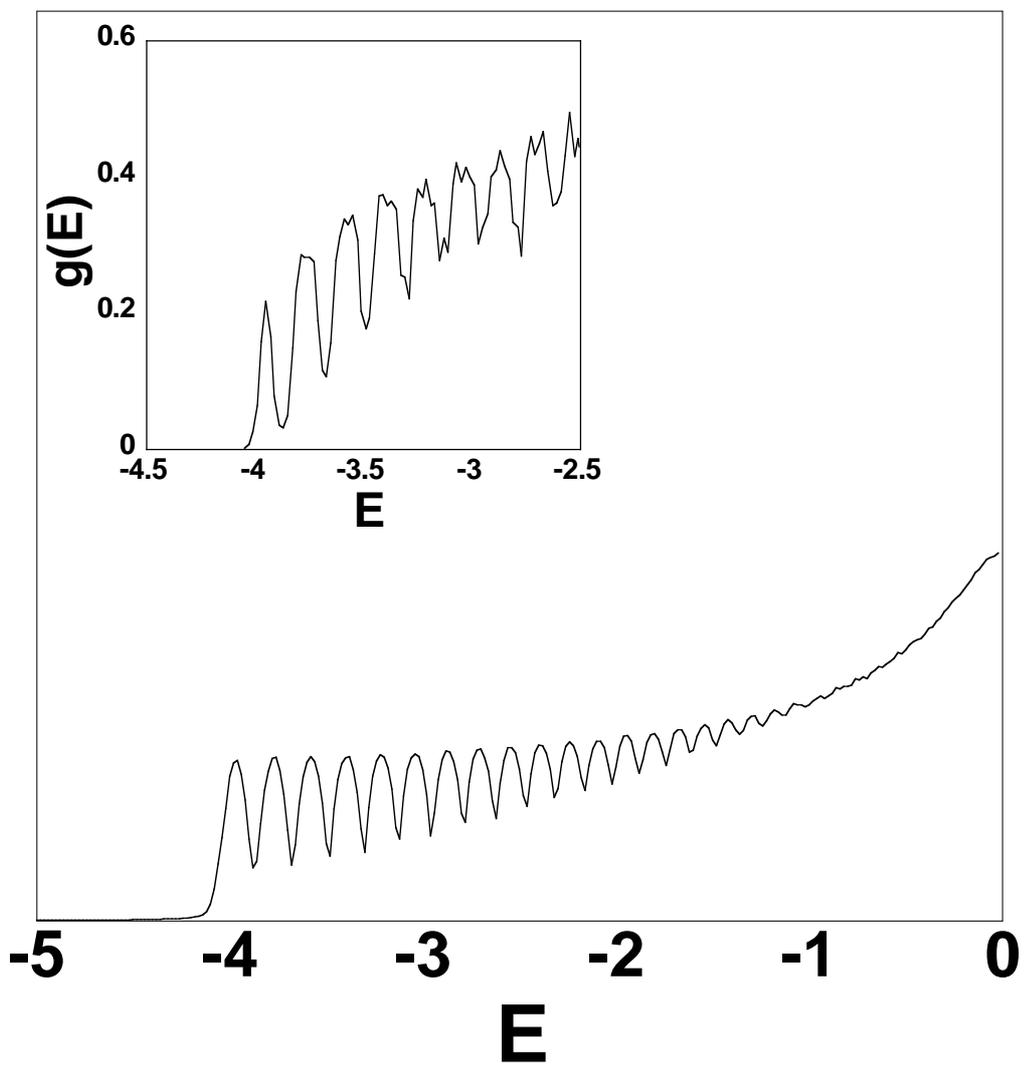
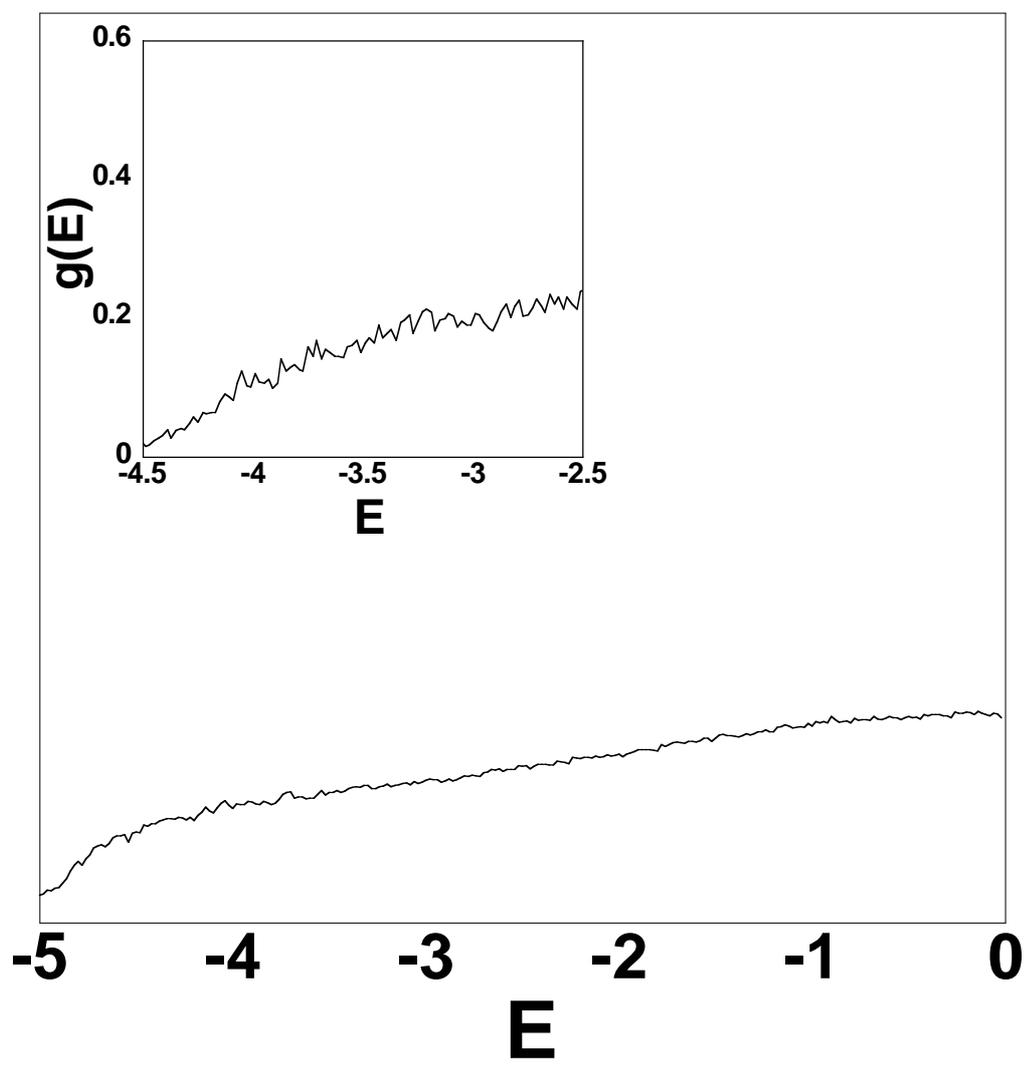
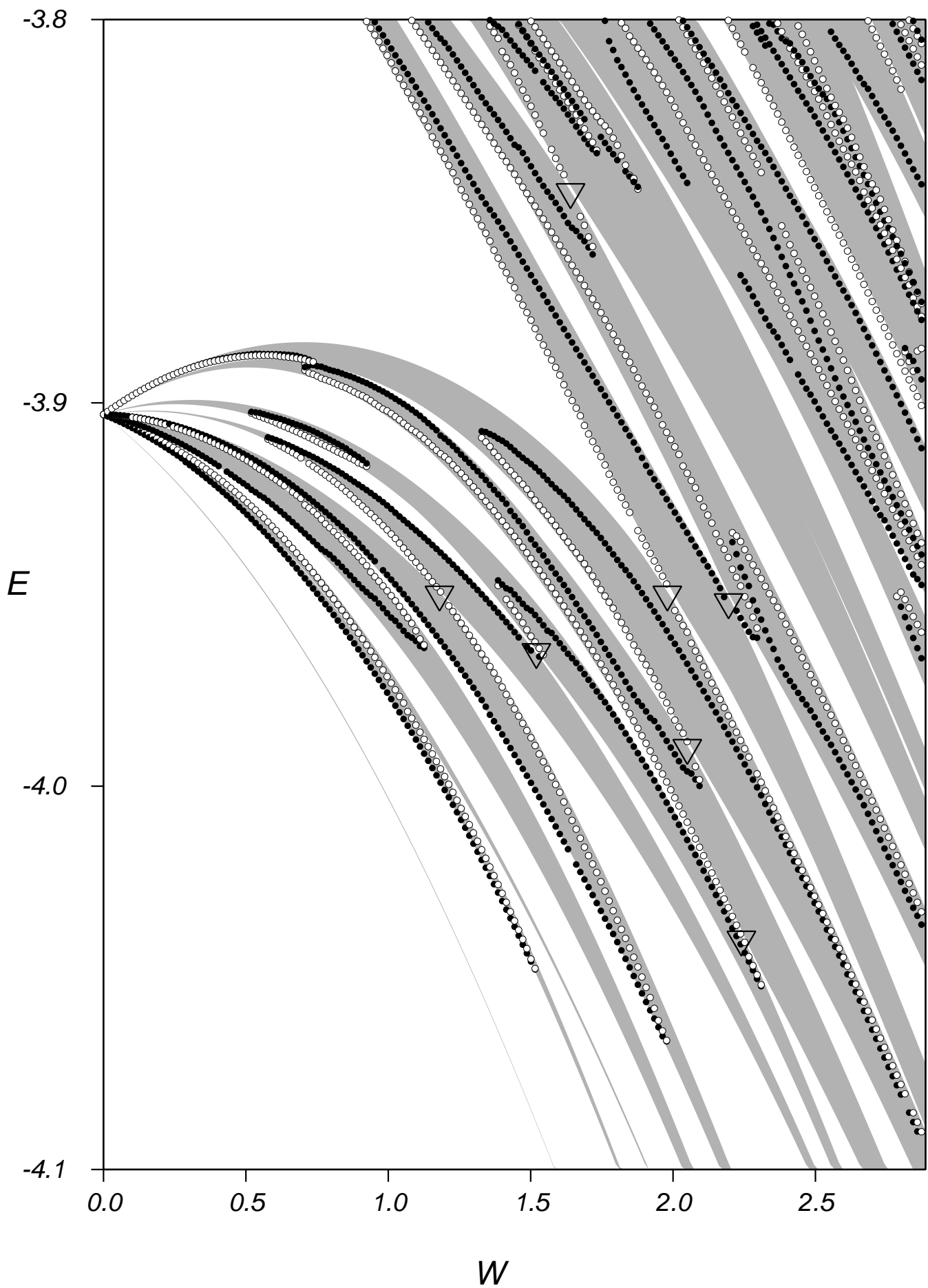
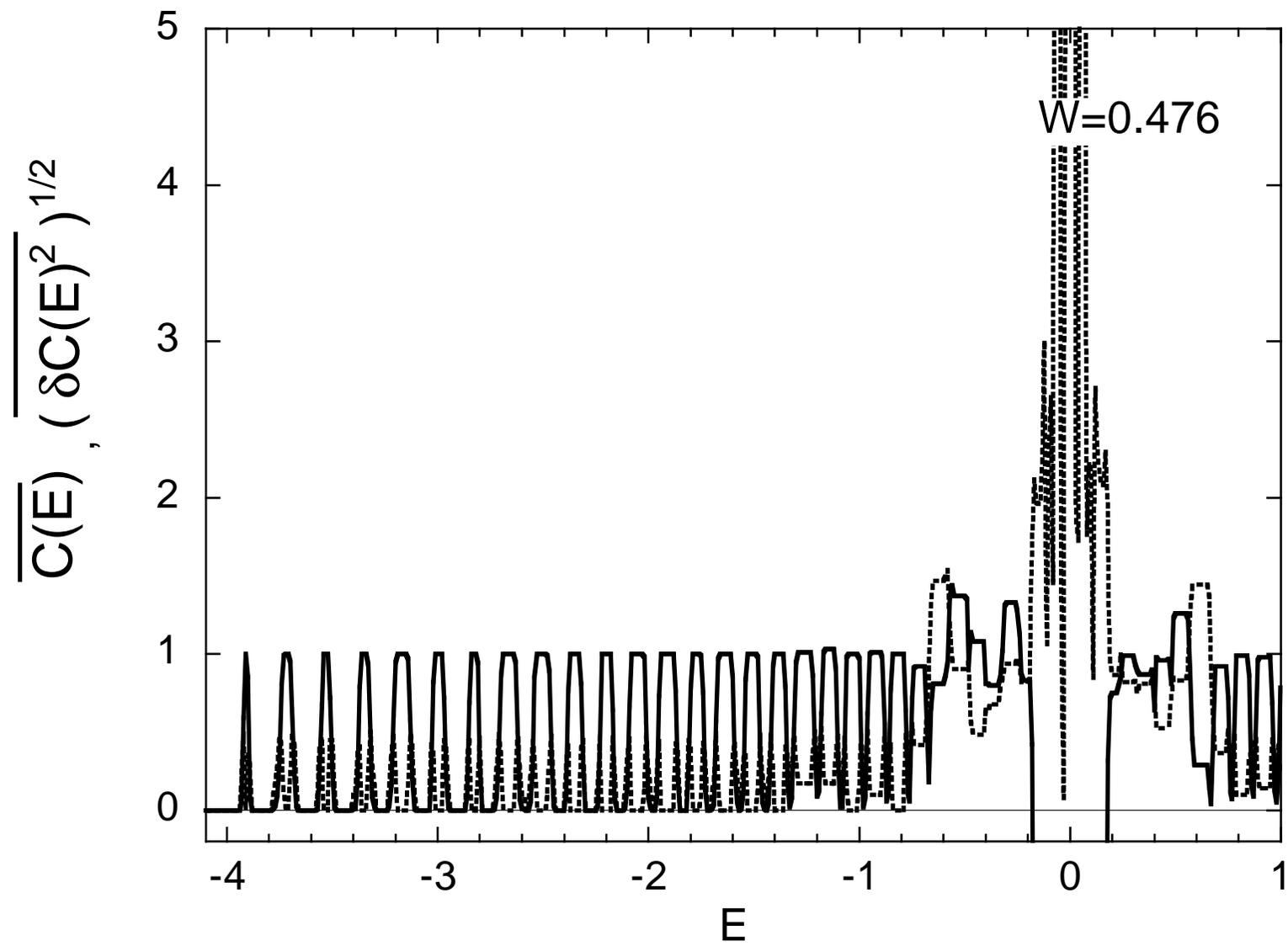
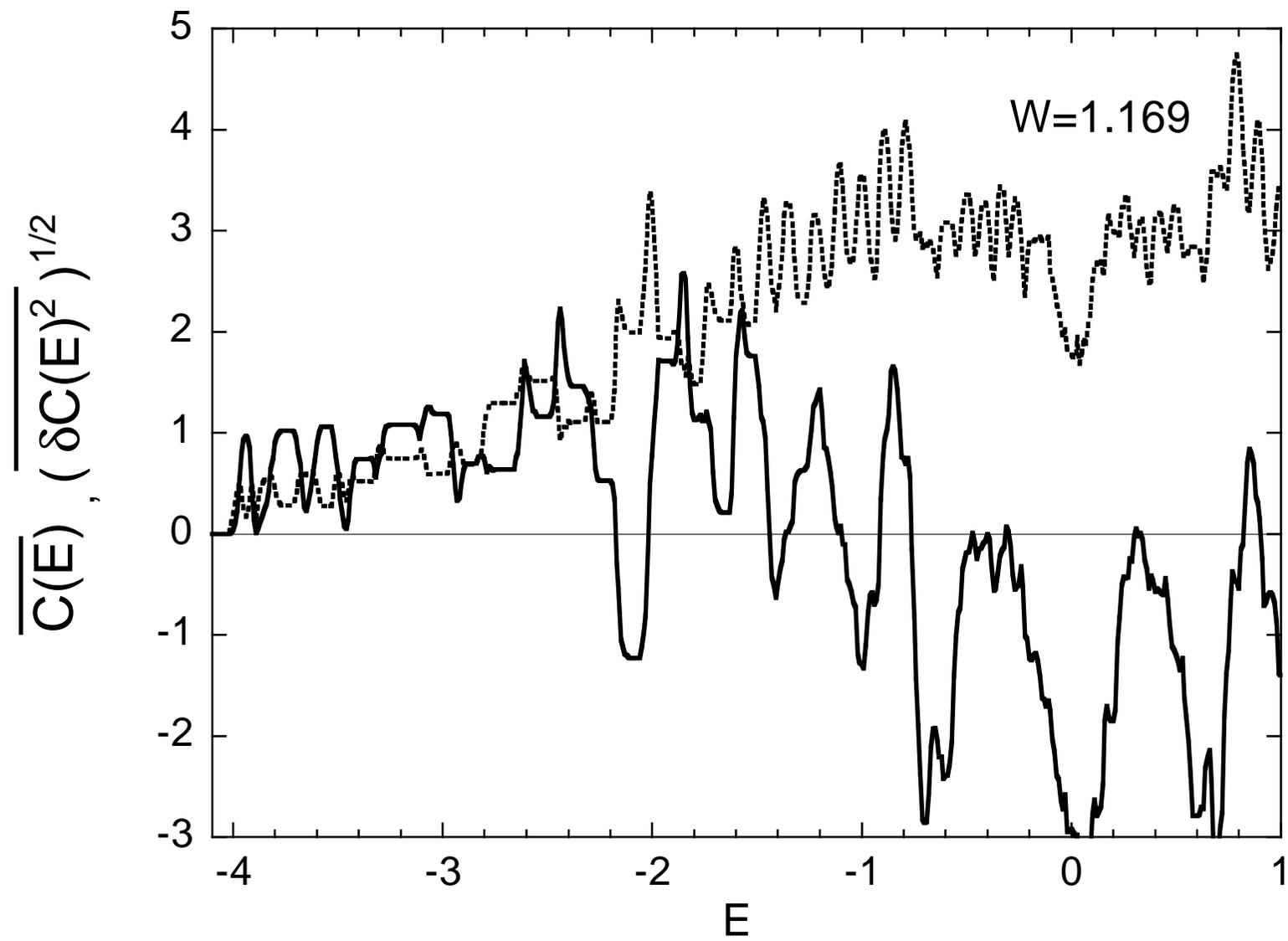
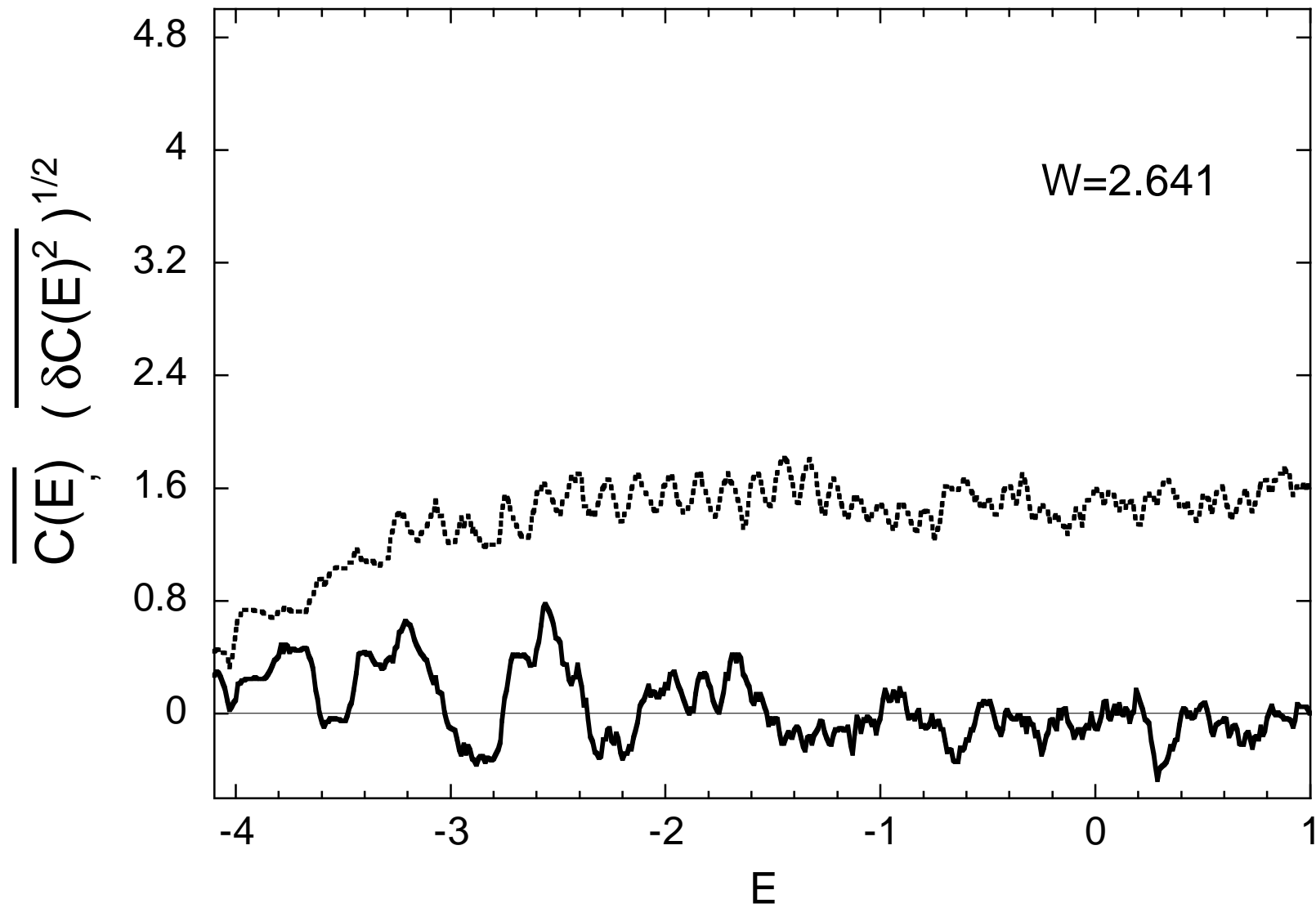
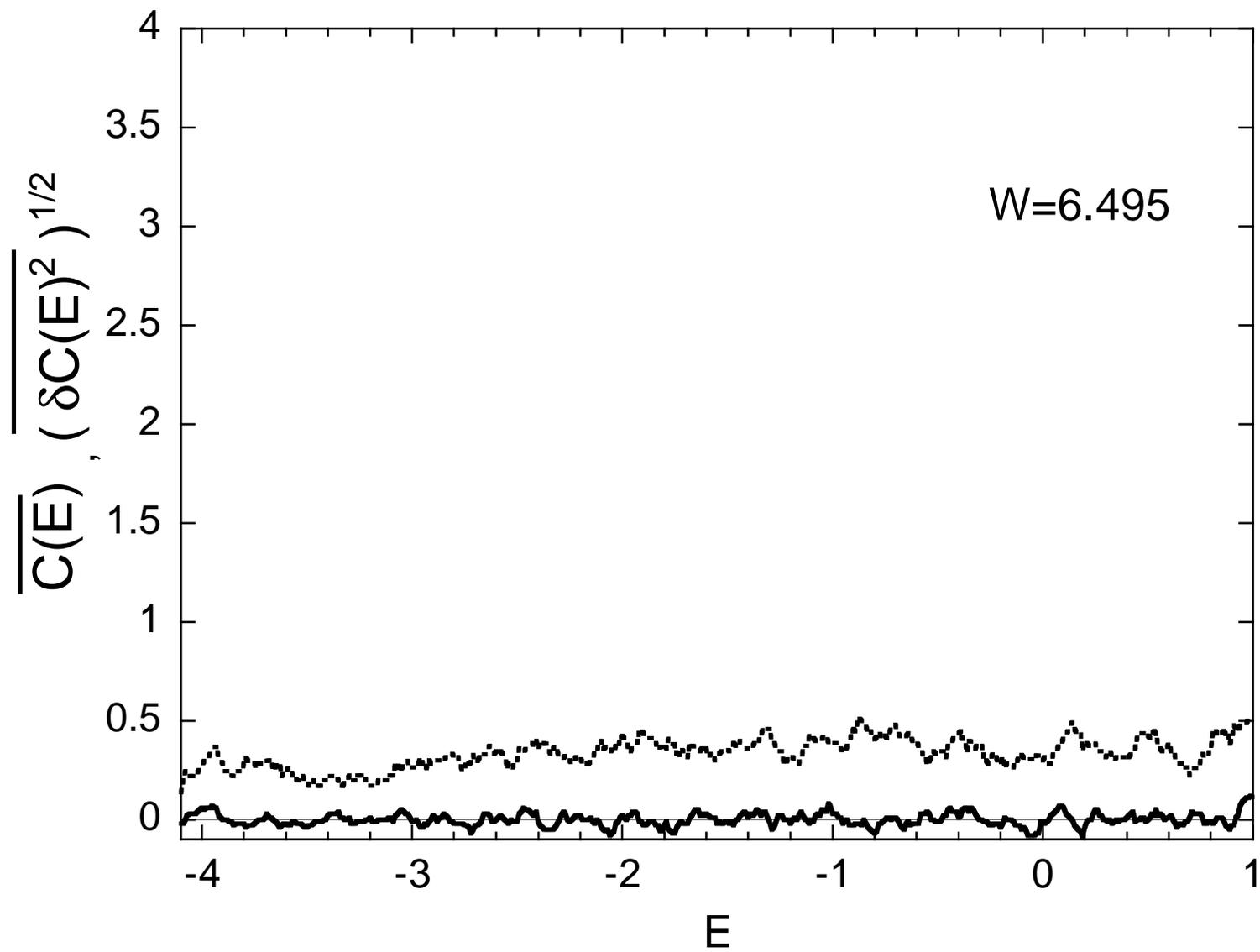
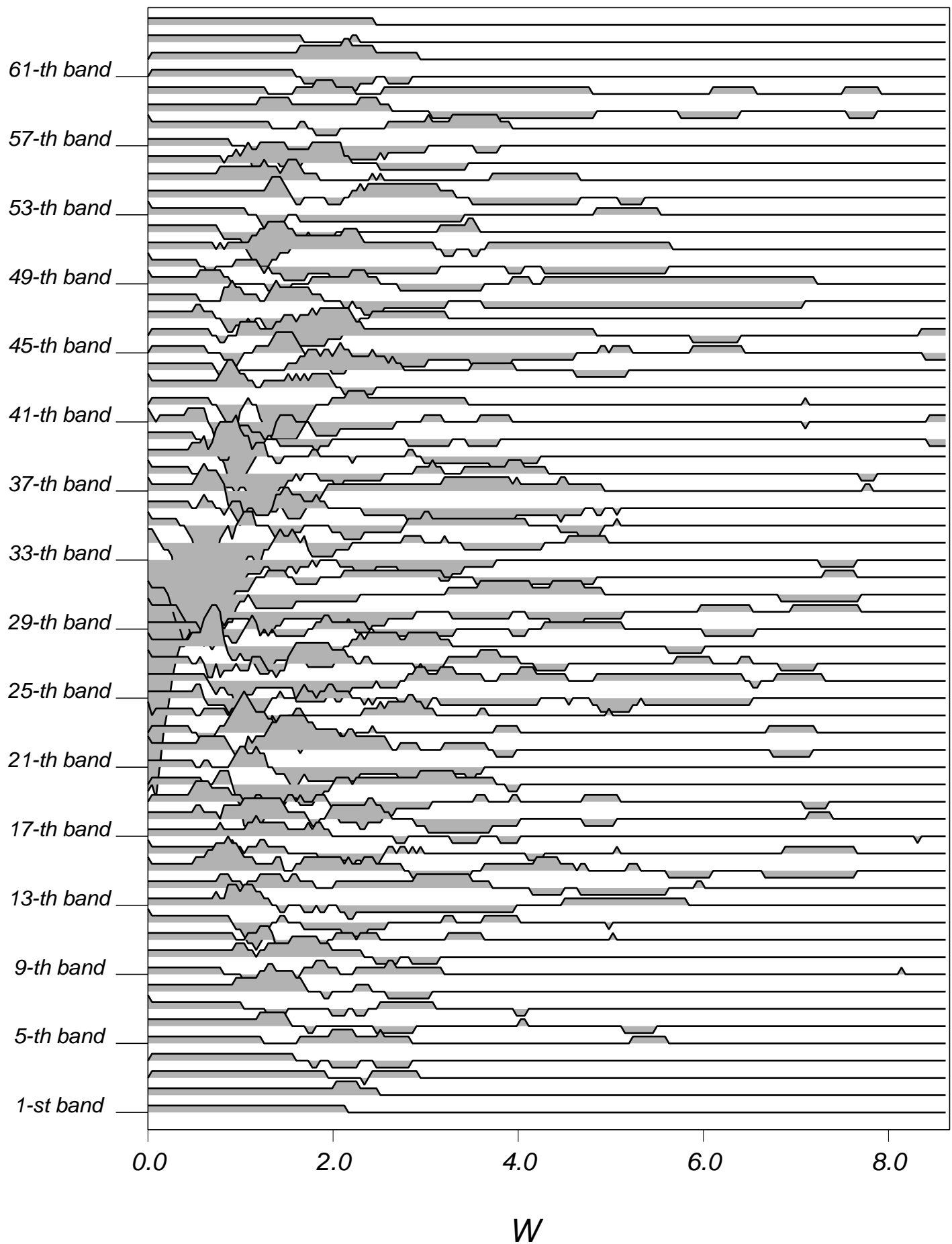
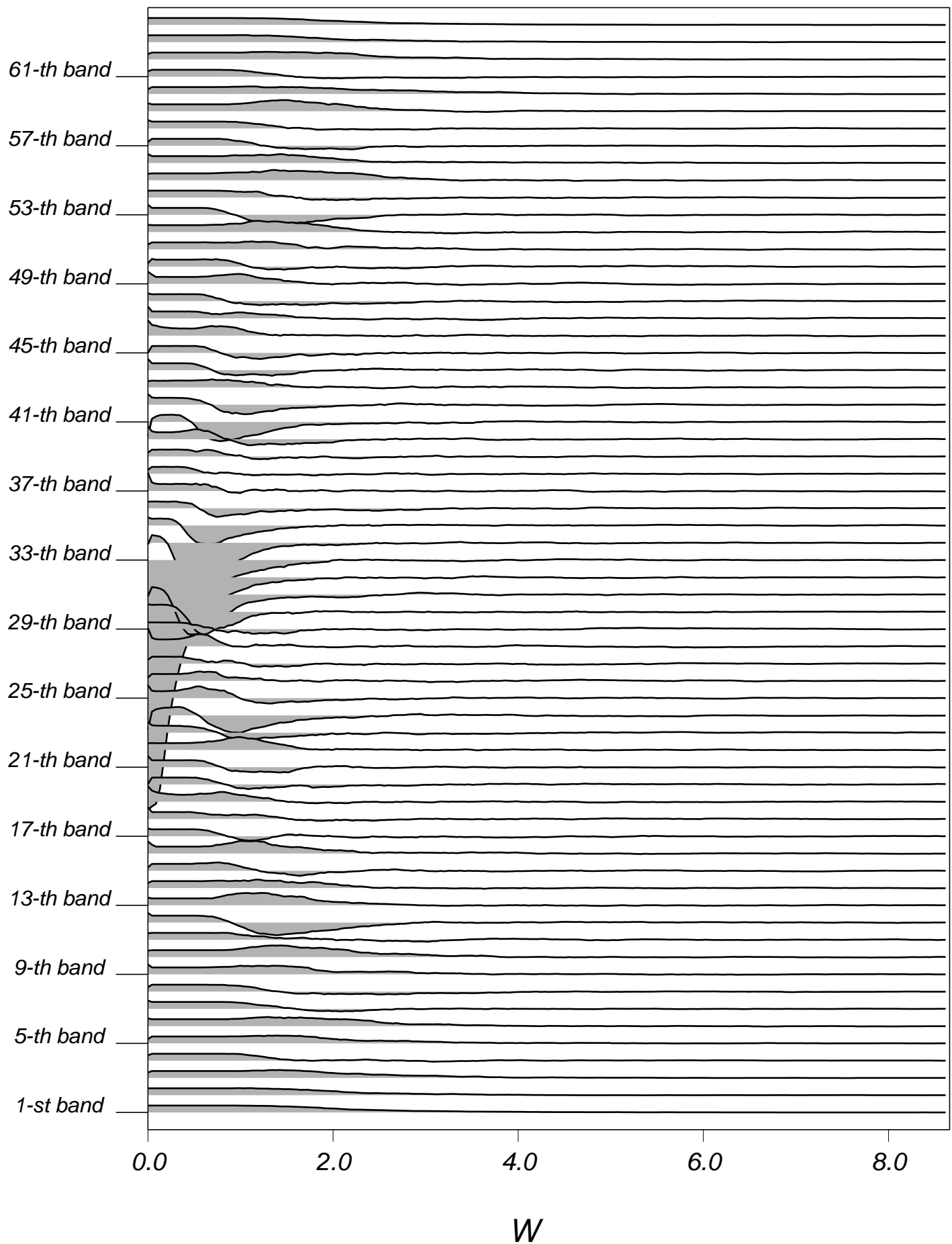
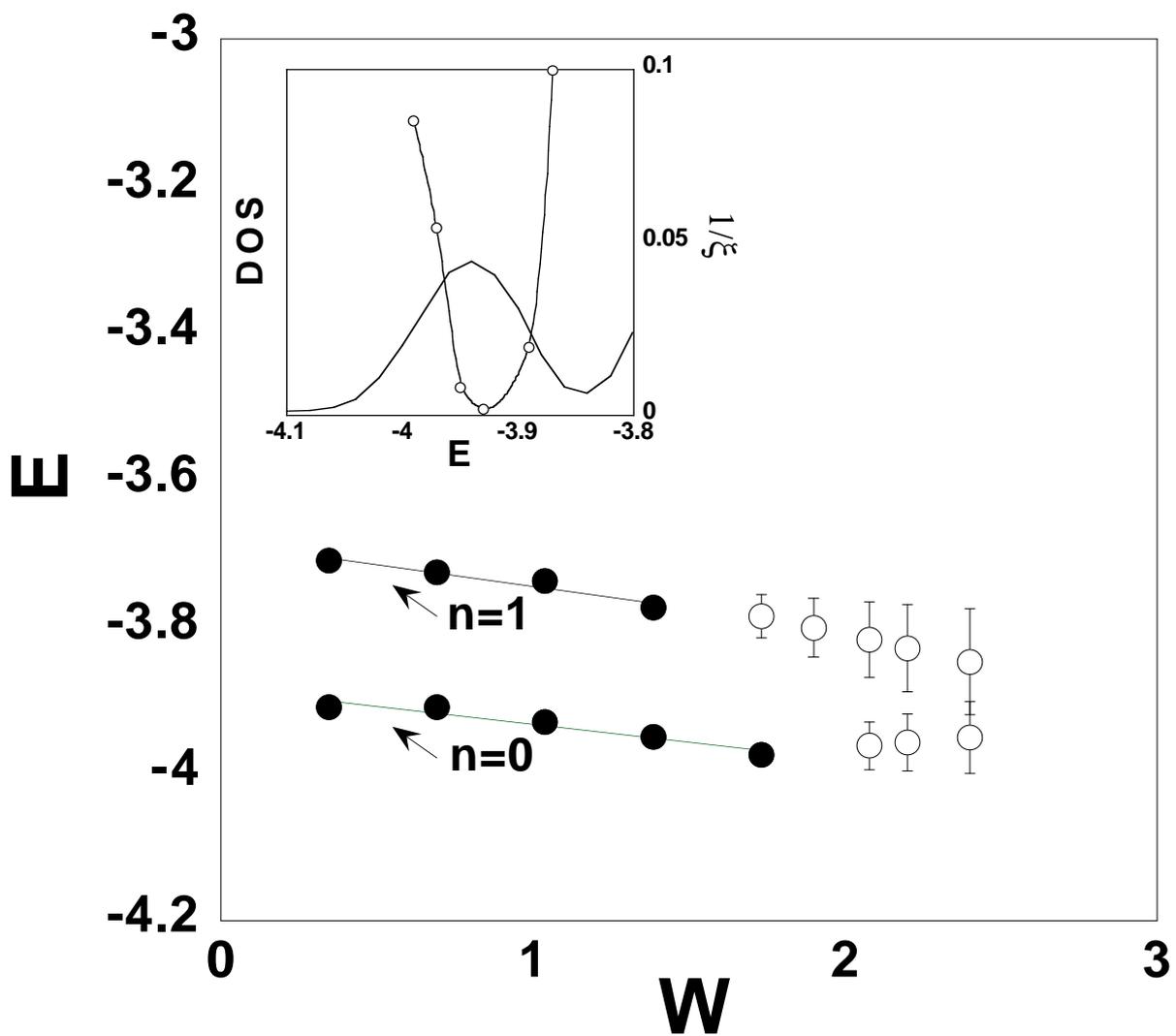
\end{document}